 \documentclass[final,5p,times,twocolumn]{elsarticle}

\usepackage{amssymb}

\journal{Applied Mathematics and Computation}

\begin{document}

\begin{frontmatter}



\title{Mathematical Modeling of Dengue Epidemic: Control Methods and Vaccination Strategies}


\author{Sylvestre Aureliano Carvalho\fntext[label3]{Departamento de F\'isica, Universidade Federal de de Vi\c cosa, 36570-000, Vi\c cosa, MG, Brazil}, Stella Ol\'\i via da Silva\fntext[label3]{Departamento de Ci\^encias Exatas, Universidade Federal de Lavras, 37200-000, Lavras, MG, Brazil},Iraziet da Cunha Charret}
\begin{abstract}
Dengue is one of the most important infectious diseases in the world, in terms 
of death and economic cost. Hence, the modeling of dengue is of great importance to help 
us understand the dynamics disease, and interfering with its spreading mathematical by the 
proposition of control methods. In this work, control strategies in an attempt to eliminate 
the \textit{Aedes aegypti} mosquito, as well as proposals for the vaccination 
campaign are evaluated. In our mathematical model, the mechanical control is accomplished through 
the environmental support capacity affected by a discrete function that represents removal of breeding. 
Chemical control is carried out from the use of insecticide and larvicidal. The efficacious of 
vaccination is studied through the removal of a fraction of individuals, proportional to the 
rate of vaccination, from the susceptible compartment and its transfer to the recovered compartment. 
Our major find is that the dengue fever epidemic is only eradicated with the use of an immunizing 
vaccine because control measures directed against the vector are not enough to halt disease 
spreading. Even where the infected mosquitoes are eliminated from the system, susceptible 
mosquitoes are still present, and infected humans cause dengue fever to reappear in the human population. 
\end{abstract}


\end{frontmatter}


\section{Introduction}
\label{I}

In recent years the field of public health has been benefit from the use of mathematical models to the study of spread epidemic of infectious disease. 
One of the first mathematical model epidemiology has been proposed by W. H. Hamer in 1906, who postulated that the development of an epidemic is 
related to the rate of contact between infected and susceptible individuals \cite{Hamer} and currently this postulate is known as the principle of mass action \cite{Wilson}.

In 1911 Ross has proposed one of the first models used for epidemiological study of malaria \cite{Ross}. In 1957 Macdonald brought a greater biological realism to  
the Ross model, improving the accuracy and interpretation of the model parameters. The contributions of Macdonald resulted in the creation of the model 
known as the Ross-Macdonald model \cite{Macdonald}, in which two differential equations represent the dynamics of the disease regardless of the age structure of 
populations and the incubation period.

In the cases of infectious diseases, situations may exist in which infected hosts once recovered again become susceptible to new infections or may become 
immune. However, there may be situations where the host contract the disease and do not acquire the disease elicited by the same agent again. In the first 
case, these diseases are usually caused by bacterial or protozoan agents \cite{Ghosh,Singh}, such as malaria for example. The second case is due to viral diseases such as 
dengue \cite{Maidana}. 

Currently, there is great interest in studying epidemic propagation, as in the case of dengue. This disease is considered the most significant in terms 
of death and economic cost, since it is responsible for more deaths and infections in humans than any other arthropod-borne viruses \cite{Gubler}. This scenario 
can be even worse, since the exact number of cases of the disease is not known. Several cases go unreported because they may clinically manifest with few 
or no symptoms \cite{Kurane}. Due to the large number of people exposed, and absence of medicines able to cure infected people, the increase of dengue epidemics 
have motivated an intensive study of mechanical and chemical control mechanisms to used on eliminate the mosquito \textit{Aedes aegypti}.

The mechanical control, relying on educational campaigns for the removal of containers that accumulate water, results in a significant decrease in 
population size of the vector when this type of control is performed on any of the seasons \cite{Arduino}. Moreover, the mechanical control is effective only upon 
the awareness of the human population to identify and eliminate outbreaks of dengue in their own homes \cite{Luz}. Chemical control, based on the use of 
insecticide and larvicide, was tested for the reduction of vector population \cite{Luz,Burattini}. More recently, in parallel, vaccination strategies have been 
proposed as vaccines. Studies have been done to develop a vaccine, evaluate and investigate its efficacy \cite{Johnson,Johansson,Julander}.

So, in this paper, we evaluate the set of these three strategies - mechanical control, chemical control, and vaccination - in order to assess the 
best way to eradicate dengue. Our study is based on the dengue epidemiological model proposed by Yang and Ferreira \cite{Yang}. We will model the effects 
by the use of mechanical control mechanisms (parameters that decrease the rate of oviposition vector, known as carrying capacity) and chemical control 
(insecticide and larvicide). Besides the mortality rates of the vector we have also included in our study the spreading of the dengue virus in humans 
and suggest vaccination strategies in a human population.

The paper is organized as follows. Section (\ref{mmfd}) introduces the ordinary differential equations of the model. In Section (\ref{cs}), mechanical and 
chemical control strategies, as well as the factors that justify the use of information based strategies in climate and rainfall modelling are described. 
It is also highlighted the effects of vaccination campaign and reported a measure of the efficiency of these mechanisms. In section (\ref{r}), the model 
solutions obtained by numerical integration are presented and the efficiency of each strategy is evaluate. Finally, in Section \ref{c} some conclusions 
are drawn.

\section{Mathematical model for dengue}
\label{mmfd}

In the study of some epidemics, it is useful to separate the vector and host populations in some classes according to their stage in the infection cycle. 
In the case of dengue, for instance, it is useful to consider the existence of three classes for the vector and four for the host. Such separation is 
suggested, because both the vector and the host contract the virus, go through a period of latency in which the individual has the virus but is not 
capable of transmitting it, became infective and, for humans, recovered from the disease \cite{Newton}.

The dengue mosquito, the \textit{Aedes aegypti} female, can be infected with the virus when feeding infected human blood. Once it have acquired the 
virus the mosquito, remains infected throughout its life. So it is feasible to divide the population of adult midges into susceptible, exposed and infected. 
Classes in turn, the human population can recover from the illness, and for that reason, in addition to the existing classes for the vector, there is the 
class of recovered humans \cite{Chan}.

The model in which the individuals are organized in these different classes from which they can entry or exit is known as 
\textit{SEIR}(susceptible - exposed - infectious - recovered) \cite{Korobeinicov}. In this type of approach, a population is represented by 
state variables (subpopulation densities) associated to the different model compartments, which envolves on time according the 
rules of interaction. Differential equations can be used to model the rules, and give every subpopulation growth rates \cite{Esteva,Andraud}.

Our mathematical model, based on the model developed by Yang and Ferreira \cite{Yang}, comprises $10$ nonlinear first order ordinary differential equations. 
These equations include variables describing two classes of individuals: humans and mosquitoes. The phase of the mosquito in life cycle are 
represented in the system $(\ref{system1})$ as follows: egg (\textit{E}), larva (\textit{L}), pupa (\textit{P}) and adult (\textit{W}), 
the latter being composed of three subpopulations differing in its stage of infection, namely, susceptible ($W_{1}$), exposed ($W_{2}$) and 
infected ($W_{3}$).

The parameters of the model are related to biotic and abiotic factors. They represent averages of values that takes into account all stages of vector 
in development, effective rates of natural mortality, and contact rates between the stages of infection of the human population and the mosquitoes.

The transition rates between the phases of egg, larva, pupa and adult are represented by the parameters $\sigma_{j}$, ($j = e, l, p, w$). In addition, 
the system $(\ref{system1})$ contains other rates: oviposition ($\phi$), mortality of eggs, larvae, pupae and adult midges $\mu_{i}$ ($i = l, p, w$); 
which are functions of temperature, humidity, and rain precipitation. Thus these rates reflect abiotic factors. However, in the present analysis, 
these functions will be replaced by, constant average values, listed in table ($\ref{tab1}$).

\begin{eqnarray}
\label{system1}
\frac{d}{dt}E(t) &=& \phi\, \left [1-\frac{E(t)}{C'} \right]\,W(t)-(\sigma_{E}+\mu_{E})\,E(t) \\
 \nonumber
\frac{d}{dt}L(t) &=&  \sigma_{e}\,E(t)-(\sigma_{l}+\mu_{l}+\mu'_{l})\,L(t)   \\
\nonumber
\frac{d}{dt}P(t) &=& \sigma_{l}\,L(t)-(\sigma_{P}+\mu_{P}+\mu'_{P})\,P(t)   \\
\nonumber
\frac{d}{dt}W_{1}(t) &=& \sigma_{P}\,P(t)-\left [\beta_{w}\frac{I(t)}{N}+\mu_{w}+ \mu'_{w}\right]\,W_{1}(t) \\
\nonumber
\frac{d}{dt}W_{2}(t) &=& \beta_{w}\,\frac{I(t)}{N}\,W_{1}(t)-(\gamma_{w}+\mu_{w}+ \mu'_{w})\,W_{2}(t) \\
 \nonumber
\frac{d}{dt}W_{3}(t) &=& \gamma_{w}\,W_{2}(t)-(\mu_{w}+\mu'_{w})\,W_{3}(t)  \\
\nonumber
\end{eqnarray}

Other parameters correspond to biotic factors, i.e. those that have direct interference of the human activity: carrying capacity, represented by the 
letter $C'$ and the chemical control represented by the function $\mu'_{i}$. The carrying capacity can be modulated through the elimination of 
breeding grounds as a result of educational campaigns to increase the population awareness, concerning the necessities to eliminate the aquatic 
population reservoirs. The chemical control is another parameter which takes into account the work of health agents responsible for preventing 
vector proliferation in every stage of its life cycle, from vector aquatic phase to adulthood. The population adult mosquitoes is composed only 
by females feeding blood.

The equations for the humans population are coupled and separated in susceptible (\textit{s}), exposed (\textit{e}), infected (\textit{i}) and 
recovered (\textit{r}) compartiments. In our model we take into 
consideration the presence of only one virus serotype. System $(\ref{system2})$ introduces a set of four first order, nonlinear ordinary 
differential equations for the human population densities.

\begin{eqnarray}
\label{system2}
\frac{d}{dt}s(t) &=& \mu_{h}-\left[\beta_{h}\,\frac{M_{I}(t)}{M(t)} + \mu_{h}\right]\,s(t)   \\ 
\nonumber
\frac{d}{dt}e(t) &=& \beta_{h}\,\frac{M_{I}(t)}{M(t)}\,s(t)-(\gamma_{h}+\mu_{h})\,e(t) \\ \nonumber
\frac{d}{dt}i(t) &=& \gamma_{h}\,e(t)-(\sigma_{h} + \mu_{h})\,i(t) \\ 
\nonumber
\frac{d}{dt}r(t) &=& \sigma_{h}\,i(t) - \mu_{h}\,r(t)  
\end{eqnarray}

\section{Control strategies}
\label{cs}
\subsection{The influence of climate and temperature}
\noindent

The climate is one of the factors that affects the environment where the \textit{Aedes aegypti} is present \cite{Nakhapakorn}. 
The \textit{Aedes} is found primarily in tropical and subtropical areas in zones having relatively constant temperatures around $24^{o}$ $C$ \cite{Setzer}. 
The temperature has direct and indirect influences on insect's development and feeding \cite{Gubler2}. Hence, the effects of temperature on the life cycle of 
insects read be modeled to understand the population dynamics of dengue vectors and can be exploited to develop appropriate vector control strategies.

In Brazil, the fall season begins in March, winter in June, the spring starts in September, and the summer in December. 
However, these four seasons are neatly marked only in the southern region, S\~ao Paulo, Mato Grosso do Sul, and in the mountain regions of 
Minas Gerais and Rio de Janeiro, occupying approximately $15\%$ of the brazilian territory. In such regions summer and winter temperatures 
are well-defined. In contrast, in the Amazonia, for example, does not exist significant variation in temperature and rainfall during the annual 
period, so basically there are no seasons. In other regions, there are only two regimes: the rainy and the dry seasons. 

Taking into consideration the climatic variability in Brazil, we have divided a year into three periods, defined in terms of the maximum and 
minimum rainfall a major determinant of the mosquito population namely, favorable, unfavorable and intermediate. Indeed, the \textit{Aedes aegypti} 
has a weak resistant at to the low  temperature. Temperatures between $22^{o}$ $C$ and $30^{o}$ $C$ are favorable in its life cycle, and the extremes, 
below $18^{o}$ $C$ and above $34^{o}$ $C$ have negative effects on their development \cite{Koppen}.

The most favorable period for the development of the vector are the months of December, January, February and March, due to high temperatures and high 
rainfall. The colder months (June, July, August and September), are unfavourable for the development of \textit{Aedes aegypti}. The intermediate period 
includes the months of April, May, October and November \cite{Souza}. The values that make up the template have daily values according to table 
($\ref{tab1}$) and were taken from studies by Yang and Luz \cite{Yang,Luz}.

  \begin{table*}[ht]
  \centering
  \begin{tabular}    {p{0.5\linewidth}p{0.125\linewidth}p{0.125\linewidth}}
 \hline
Meaning of the Parameters & Symbol & Value (day$^{-1}$) \\
\hline
Oviposition & $\phi$ & $1$ \\
Transformation from egg to larva(F) & $\sigma_{e}$ & $0.330$ \\
Transformation from egg to larva (U) & $\sigma_{e}$ & $0.300$ \\
Transformation from egg to larva (I) & $\sigma_{e}$ & $0.200$\\
Transformation from larva to pupa (F) & $\sigma_{l}$ & $0.140$\\
Transformation from larva to pupa (U) & $\sigma_{l}$ & $0.125$\\
Transformation from larva to pupa (I) & $\sigma_{l}$ & $0.066$\\
Transformation from pupa to mosquito (F) & $\sigma_{p}$ & $0.346$\\
Transformation from pupa to mosquito (U) & $\sigma_{p}$ & $0.323$\\
Transformation from pupa to mosquito (I) & $\sigma_{p}$ & $0.091$\\
Mortality rates of egg & $\mu_{e}$ & $0.050$\\ 
Mortality rates of larva & $\mu_{l}$ & $0.050$\\ 
Mortality rates of pupa & $\mu_{p}$ & $0.0167$\\
Mortality rates of mosquito (F) & $\mu_{w}$ & $0.042$\\
Mortality rates of mosquito (U) & $\mu_{w}$ & $0.040$\\
Mortality rates of mosquito (I) & $\mu_{w}$ & $0.059$\\
Force of infection human infectious$\rightarrow$mosquito susceptible & $\beta_{w}$ & $0.750$\\
Force of infection human susceptible$\rightarrow$mosquito infectious & $\beta_{h}$ & $0.375$\\
Incubation period of virus (mosquito) & $\gamma_{w}$ & $0.200$\\
Incubation period of virus (in humans) & $\gamma_{h}$ & $0.100$\\
Infectious (or recovery) period (in humans)   & $\sigma_{h}$ & $0.143$\\
Rates per-capita mortality human & $\mu_{h}$ & $0.000042$\\
Rates per-capita natality human & $\mu_{n}$ & $0.00042$\\
\hline
  \end{tabular}
  \caption{The parameters and their values used in the model. The transformation rates in the aquatic phase as well as rate mosquitoes mortality are subdivided into periods: Favorable (\textit{F}), Unfavorable (\textit{U}) and intermediate (\textit{I}). The values have units of day$^{-1}$.}
  \label{tab1}
\end{table*}
   
\subsection{Mechanical Control}
\label{mc}

It is widely known that a population can not grow indefinitely because the environment has a limited capacity of resources to sustain their individuals. 
This ability is known as environmental supportability \cite{Gotelli}. There population size of the dengue mosquito increases considerably  rainy 
season due to the increase in the number of containers with standing water, ideal places for mosquito oviposition \cite{WHO1}. Thus, 
the environmental support capacity of the dengue vector is primarily related to the amount of available sites for laying eggs, since the resources necessary 
for the maintenance of adults vectors such, as food and shelter, are virtually unlimited.

In the model, described by equations $(\ref{system1})$, the environmental support capacity is given by $C'$ for the population of eggs, and is interpreted as the 
physical space available for mosquitoes oviposition. A method to control the growth of mosquito population is to decrease $C'$ through the elimination 
of breeding grounds. This is usually done mechanically, at different times of the year, accordingly the climatic conditions.

In order to take ino account this control strategy, our model assumes that the support capacity $C'$ is a function of a random variable $C_{i}$. 
The effectiveness reate of the control performed at given moment. Specifically,
\begin{eqnarray}
\label{system3}
C'&=& C_{i}C_{fixo}
\end{eqnarray}
where $C_{fixo}$ is the supportability of the containers in the absence of control mesures and
\begin{eqnarray}
0 < C_{i} \leq 1
\end{eqnarray}
varies according to the time of year. 

When $C_{i}$ receives values close to $0$, leading to a very small $C'$, there are few locations for oviposition. In contrast, when  
$C_{i}\sim1$, the support capacity remains virtually unchanged, and there is no reduction of mosquito breeding sites. In turn, the value of $C_{fixo}$  varies 
with the annual seasons.

Within the favorable period, characterized by high temperatures and humidity \cite{Camara}, a value $C_{fixo}= 700$ was fixed. Within (cold and arid) the 
unfavorable period \cite{Koopman} we set $C_{fixo} = 300$. For the months that comprise the mild period, a supportability $C_{fixo} = 500$ is assumed.

\subsection{Chemical Control}

There are two types of chemical that hill the dengue vector: insecticides, that affect adult mosquitoes, and larvicides, that kill vector larvae. 
We assume that the chemical control does not act on the population of pupae ($\mu'_{p}=0$). 
The insecticides is the most used instruments for vector control. 
In turn, larvicides are complementarly imployed to this control. Hence, in our model, these chemical methods are represented by the variable 
$\mu'_{w}$ an $\mu'_{l}$ that affect winged mosquitoes and larvae, respectively. These mortality rates are exponential decay functions. 
They change with the period of the year as shown in table ($\ref{tab4}$). It is assumed that the chemicals are released during the first weeks of every month, 
and persists for $15$ days in the environment. Also, at the first day the chemical release is the maximum, and at most $10\%$ of the total amount 
of released chemicals remain in the environment.

  \begin{table*}[ht]
  \centering
  \begin{tabular}    {p{0.25\linewidth}p{0.25\linewidth}p{0.25\linewidth}}
 \hline
Period & Insecticide function & Larvicide function \\
\hline
Unfavorable & $\mu^{'}_{w}(t)=0.941\,e^{-0.00149t}$ & $\mu^{'}_{l}(t)=0.884\,e^{-0.00145t}$\\ 
Intermediate & $\mu^{'}_{w}(t)=0.960\,e^{-0.00151t}$& $\mu^{'}_{l}(t)=0.825\,e^{-0.00141t}$\\
Favorable & $ \mu^{'}_{w}(t)=0.958\,e^{-0.00151t} $ & $ \mu^{'}_{l}(t)=0.810\,e^{-0.00139t} $ \\
\hline
  \end{tabular}
  \caption{Continuous functions representing the control by insecticide and larvicidal, at the favorable, unfavorable and intermediate periods.}
  \label{tab4}
\end{table*}

\subsection{Vaccination Campaign}

Currently the fight against dengue is restricted to eradication methods of its vector. The development of new tools to combat dengue, as for example, 
the discovery of a vaccine requires more knowledge about the biological characteristics of the virus and its interaction with the hosts. 

Our model, described by equations $\ref{system1}$ and $\ref{system2}$, does not take into account both people who could die 
by the disease and does not describe the existence of a vaccine that makes humans immune to dengue virus. 

The development of a tetravalent vaccine against the dengue virus is a promise of the World Health Organization (WHO), but efforts to develop vaccines have 
been hampered by the lack of an appropriate animal test subject \cite{WHO2}.

In order to take into account these two features of the epidemics, we extended our model as follows. The death of infected human introduced through the 
$\mu_{d} = 0.04$ addition of a mortality rate in the infected compartment:
\begin{equation}
\label{eqI}
 \frac{d}{dt}i(t) = \gamma_{h}\,e(t)-(\sigma_{h} + \mu_{h} + \mu_{d})\,i(t) 
\end{equation}

To represent a possible vaccination campaign, a fraction proportional to a vaccination rate $\nu = 0.2$ of susceptible individuals is removed from  
and added directly to the recovered compartment. Also, the efficiency of the vaccine is takes into account through a variable 
$\xi$. If $\xi=0$, the vaccine is not effective and does not change the 
system, whereas, if $\xi=1$, the vaccine is fully effective, and all the vaccined individuals became immunized. So, the effect of a vaccination campaing on 
the susceptible and recovered populations are given by:
\begin{eqnarray}
 \frac{d}{dt}s(t) &=& \mu_{h} - \left[\beta_{h}\,\frac{M_{I}(t)}{M(t)} + \mu_{h} \right]\,s(t) -\xi\,\nu\,s(t)\label{vac} \\
 \frac{d}{dt}r(t) &=& \sigma_{h}\,i(t) - \mu_{h}\,r(t) + \nu\,\xi\,s(t)
\label{vac1}
\end{eqnarray}

The duration of the campaign is varied to find out the time needed to eliminate the virus using a vaccination campaign. 

\subsection{Efficiency}

For each control methods, its percentual efficiency $\Sigma$ can be defined as
\begin{equation}
\Sigma = \left(1-\frac{A_c}{A_s} \right )100\%
\label{efic}
\end{equation}
where $A_{C}$ represents the area under the affected mosquito curve population as a function of time when the controlmethod is used and $A_{S}$ 
is the area under the affected total population curve without the use of the method.

\section{\textbf{Results}}
\label{r}

The results refer to the numerical simulation for a time period of approximately $20$ to $50$ years. The parameter values used by the table ($\ref{tab1}$) were adopted in order to describe coexistence between populations of humans and mosquitoes (aquatic and adult). Initially, we consider that the human population grows exponentially, as suggested by the United Nations (UN). Finally, simulation for a stationary human population is considered.

\subsection{Human population with exponential growth}

Dengue is a disease recognized in more than $100$ countries around the world \cite{Bricks}. According to the UN, over $70\%$ of the world's population live in underdeveloped countries. Asia, with a population of $4.1$ billion inhabitants, and Africa, with $1.031$ billion inhabitants exhibit, both a high population growth rate, which can be characterized as exponential \cite{PRB}. Since in those continents there is the presence of dengue viruses, there is much current interest in analyse the dengue epidemics in such exponentially growing populations.

\subsubsection{Mechanical control}

The Aedes aegypti female lays its eggs in places containing stagnant water which nurtures the development of immature stages of the vector. Consequently, if breeding grounds are eliminated, a decrease in the population size of the vector and, consequently, in cases of dengue fever will expected. In order to test the impact of the mechanical control on the population of adult vectors, a periodical decrease has been made in the carrying capacity, following the rule described in section ($\ref{mc}$).

\begin{figure}[htb!]
 \centering
 \includegraphics[width=8.5cm]{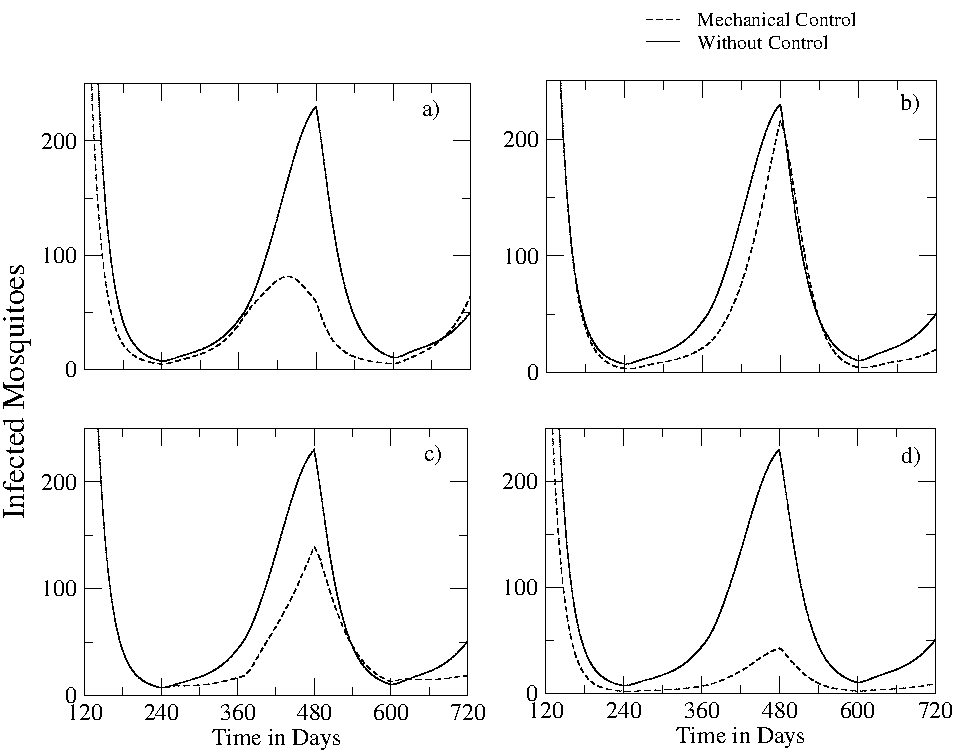}
 \caption{Comparison between the population sizes of the dengue vector with and without mechanical control. The control is held only in (a) the most favorable period, (b) unfavorable period, (c) intermediate period and (d) throughout the year. The case when mechanical control is performed (deshed line) is compared to the case in which no control type was performed (full line).}
 \label{fig1}
\end{figure}

\begin{figure}[htb!]
 \centering
 \includegraphics[width=8.5cm]{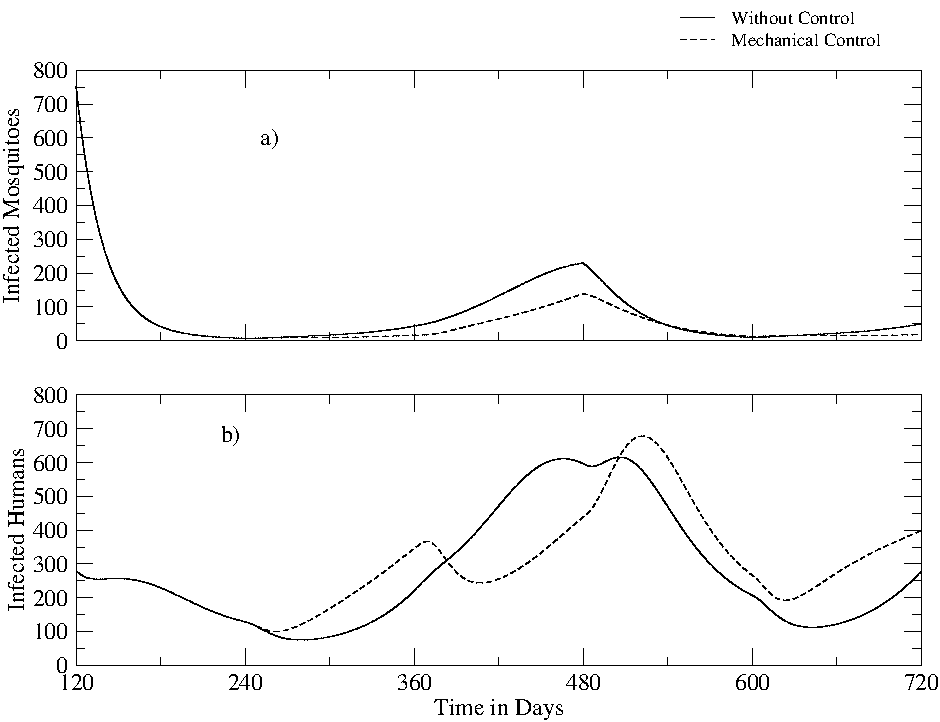}
 \caption{(a)Population size of mosquitoes infected with and without the mechanical control in the intermediate period. (b) Population of humans infected when no control is performed and when the control is performed in the intermediate period.}
 \label{fig2}
\end{figure}

Figure ($\ref{fig1}$) compares the results when the mechanical control was accomplished distinct over annual periods (favorable, unfavorable, intermediary and throughout the year) (deshed line) with those in which no control was performed (full line). The greatest efficiency was obtained when the mechanical control was carried out throughout the year, as shown ($\ref{fig1}-d$). Figure ($\ref{fig2}$) describes the evolution of the infected mosquito and human populations, when no control is performed (full line) and when the mechanical control is performed only in the intermediate period (dashed line). In the time interval between $240$ and $360$ days, there was an increase in the number of disease case when compared to simulations in which there were no mechanical control. However, in the next interval (between $360$ and $480$ days) this number was smaller. This variation in the number of cases is due to the interference of mechanical control. Reflexes are observed in the period after the intervention, since this type of control eliminates the aquatic stages of the mosquito.

When the control is performed only in one of the annual periods, an increase in the number of humans infected in period, even with a reduction in the number of infected vectors as shown in figure $\ref{fig2}$.

We can realize greater efficiencies through the use of mechanical control throughout the year. However, the \textit{Aedes aegypti} mosquito is not eradicated only using the mechanical control and, consequently, the virus remains in the human population. Hence, the use of insecticide and larvicidal was tested in the following section in an attempt to exterminate the disease. The corresponding results are discussed below.

\subsubsection{Chemical Control}

Christophers found that the mosquito in larval and adult stages are billed by of chemicals applied during their development \cite{Christophers}. We use the functions in the table ($\ref{tab4}$), in order to evaluate the impact of insecticides and larvicidaes on the control of dengue epidemic. In addition, to discriminate the impact of different types of chemical control, simulations were performed considering only one type of control at a time.

\begin{figure}[htb!]
 \centering
 \includegraphics[width=8.5cm]{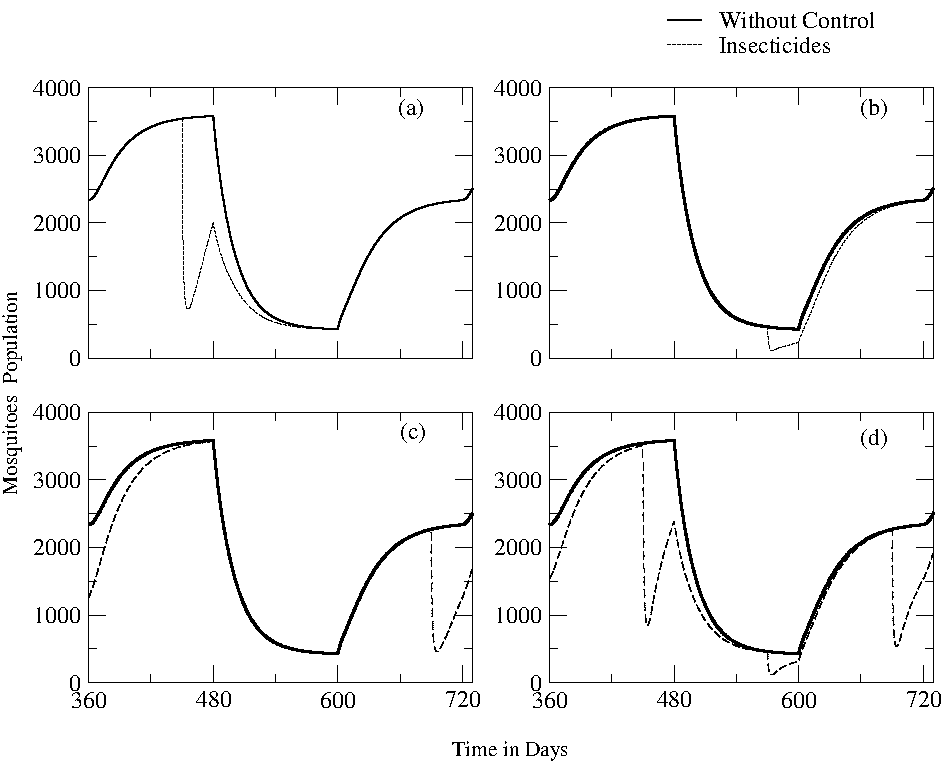}
 \caption{Mosquito population sizes after applying the insecticide only in (a) the  most favorable, (b) unfavorable, (c) intermediate period, and (d) throughout the year.}
 \label{fig3}
\end{figure}

Figure ($\ref{fig3}$) presents the population size of the \textit{Aedes aegypti} mosquito when an insecticide was applied. As can be seem, the use of the insecticide was not effective in eliminating the mosquito population, and when the application was made throughout the year (Figure 3-d), the impact on the total population was greater. However this control had only $21\%$ efficiency, as calculated according to the equation $\ref{efic}$.

\begin{figure}[htb!]
 \centering
 \includegraphics[width=8.5cm]{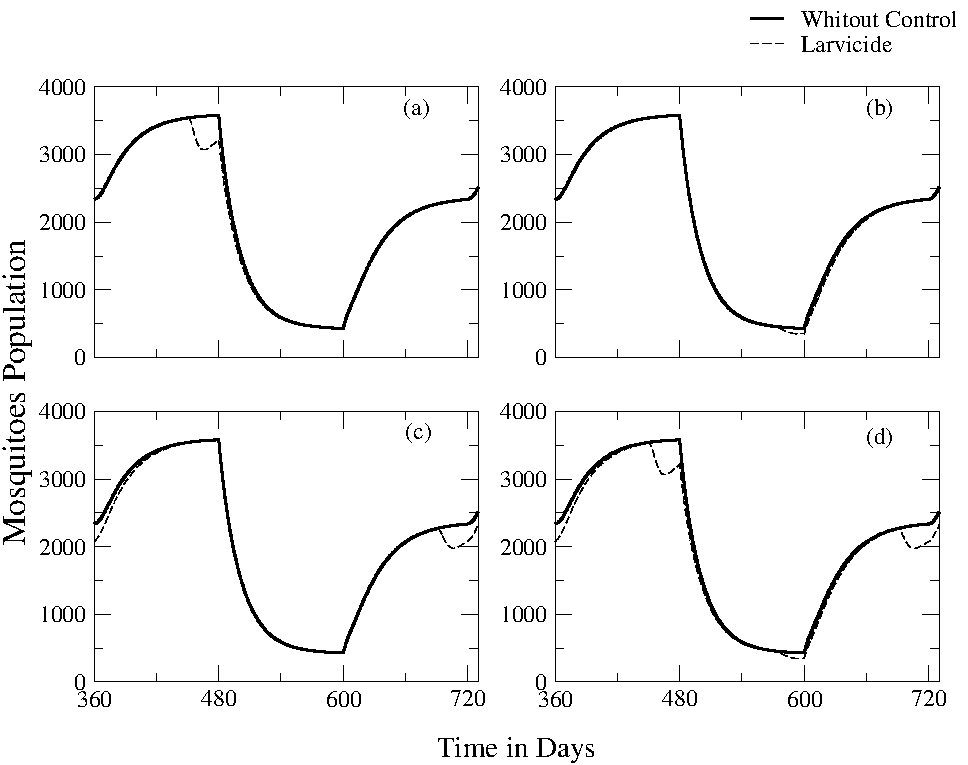}
 \caption{Comparison of the population size of mosquitoes after application of larvicidal when the control is held (a) only in the most favorable period, (b) the unfavorable period, (c) the intermediate period and (d) throughout the year.}
 \label{fig4}
\end{figure}

In ($\ref{fig4}$) are shown the results for the application of larvicidal. The approximate efficiency is $5\%$, and the population practically evolves as if no control mechanism was used. This indicates that such control is not feasible in an attempt to eradicate the \textit{Aedes aegypti} mosquito.

\begin{figure}[htb!]
 \centering
 \includegraphics[width=8.5cm]{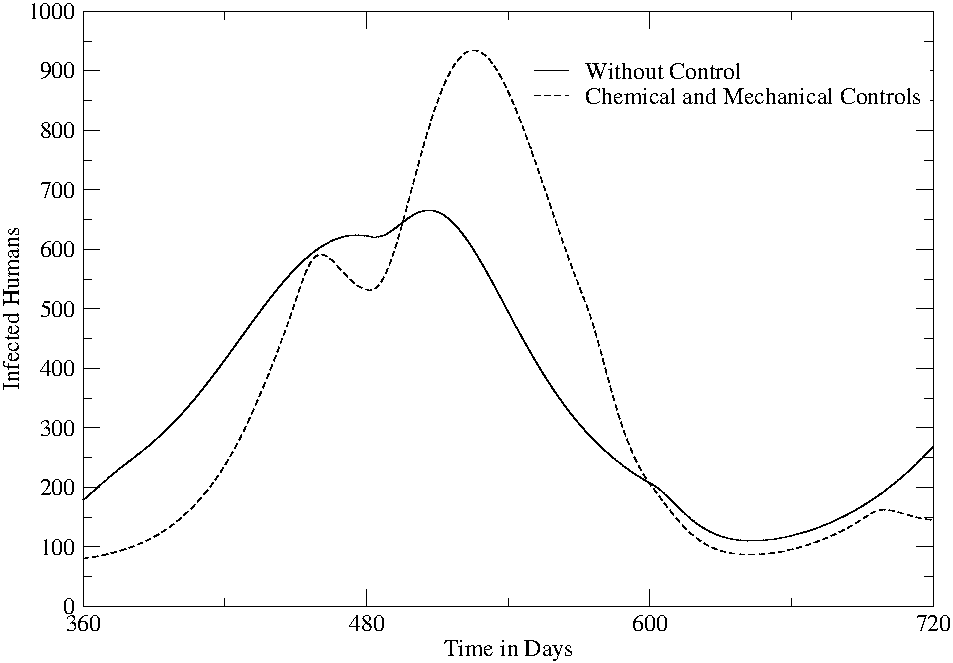}
 \caption{Comparison of the population size of humans infected when no control is carried out and when the chemical and mechanical control are performed together throughout the year.}
 \label{fig5}
\end{figure}

Figure ($\ref{fig5}$) shows the effect of chemical control on the population of infected humans. The result is independent on the chemical applications. 
Even with a decrease in the number of individuals at a given time, this number increases again latter leading do a control efficiency of less than $1\%$. 
This may indicate that the density of infected mosquitoes does not interfere with the virus permanence into the system, but simply there is a critical population 
size that guarantees virus support. Accordingly, in countries in an endemic state, most of their districts presented a vector index below the value required 
for the risk of dengue fever epidemics \cite{Camara}. The incidence of infection was high even in areas with low numbers of positive dengue mosquito breeding sites 
\cite{Teixeira}. Furthermore, small breeding are as with mosquito outbreaks in their neighborhoods can ensure the presence of dengue virus in all other districts of 
a city \cite{Braga}.

The long-term control done by chemical, such as those considered here, should be analysed with caution because its continued use can generate resistant vectors populations \cite{Braga}. 
In addition, according to the technical note $n^{o}109/2010$ \cite{MDS}, it is important to consider that there are no insecticides suitable for use in public health, since all these 
products are used specifically for agriculture.

\subsubsection{Mechanical and chemical control}

Since the use of insecticide and larvicidal separately was inefficient to eliminate the population of dengue adult vectors, we considered a scenario in which the two chemicals were applied simultaneously with the mechanical control throughout the year. 

\begin{figure}[htb!]
 \centering
 \includegraphics[width=8.5cm]{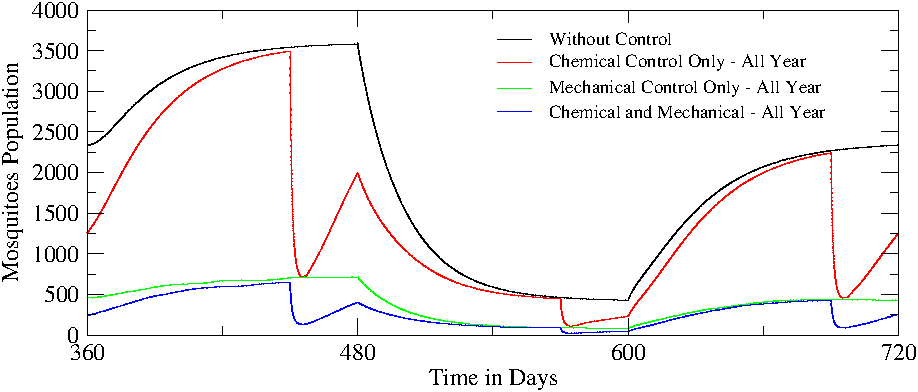}
 \caption{(Color online) Comparison between the different applications of chemical and mechanical controls. The combined use of chemical mechanical controls resulted in an $86\%$ efficiency (blue curve). In turn, the efficiency fell to $26\%$ it only the mechanical control (green curve) is used.}
 \label{fig6}
\end{figure}

The results are shown in fig. ($\ref{fig6}$) for different control strategies. When chemical control was combined with the mechanical control, the efficiency is $86\%$ (blue curve). In comparison, if only the mechanical control (green curve) was used the efficiency fell to $26\%$. However, even if both types of control are used, it has not been possible to eliminate dengue from the system, still leaving the human population at risk of contracting the disease.

\subsubsection{Death by disease and vaccination campaign}

In previous simulations the fact that people can die due disease complications was not considered. 
Dengue fever, especially in its most severe form, is hemorrhagic, and may have a lethality greater than $20\%$ \cite{Bricks}. 
Hence, in order to increase the model's  realism, we add to the human infected compartment, eq. ($\ref{eqI}$), the rate of death by illness $\mu_{d}$. 
Since, the Epidemiological Bulletin of the Ministry of health in the year of $2005$ reported a $4\%$ lethality level for dengue, we adapt $\mu_{d} = 0.04$ \cite{MDS2}.

\begin{figure}[htb!]
 \centering
 \includegraphics[width=8.5cm]{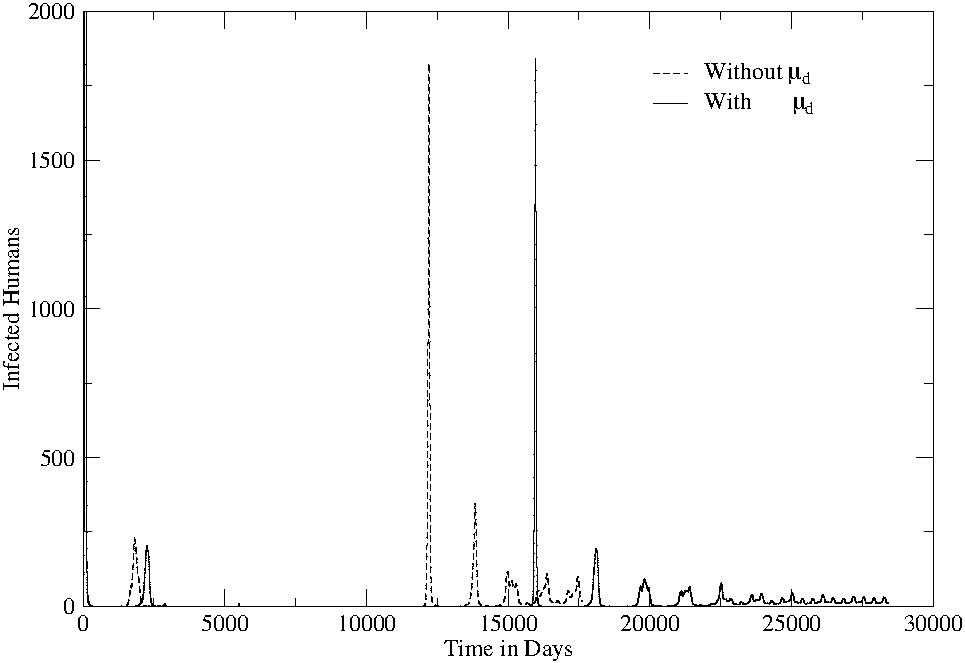}
 \caption{Evolution in time of the human infected population with and without the disease mortality rate parameter $\mu_{d}$.}
 \label{fig7}
\end{figure}

This new factor has a quantitative impact on disease outbreaks, leading to outbreaks delays at long-term , as can be seen in figure ($\ref{fig7}$). 

Because, the chemical an mechanical controls were unable to eliminate the dengue virus circulation in the human population, we tested the effects of an eventual vaccine against dengue fever on its spreading dynamics. In order, to simulate a vaccination, a fraction, proportional to the vaccination rate, $\nu = 0.2$, is withdrawn from the compartment of susceptibles, equation ($\ref{vac}$), and directly added into the recovered compartment, equation ($\ref{vac1}$). 

It is widely known that vaccines licensed by the health agencies are not $100\%$ effective. Due to this fact, the efficiency of the vaccine was also considered and represented by a variable $\xi$. $\xi = 0$ corresponds to an ineffective vaccine that does not develop immunity, whereas $\xi = 1$ means that the vaccine is fully effective, immunizing all the individuals received the vaccine.

\begin{figure}[htb!]
 \centering
 \includegraphics[width=8.5cm]{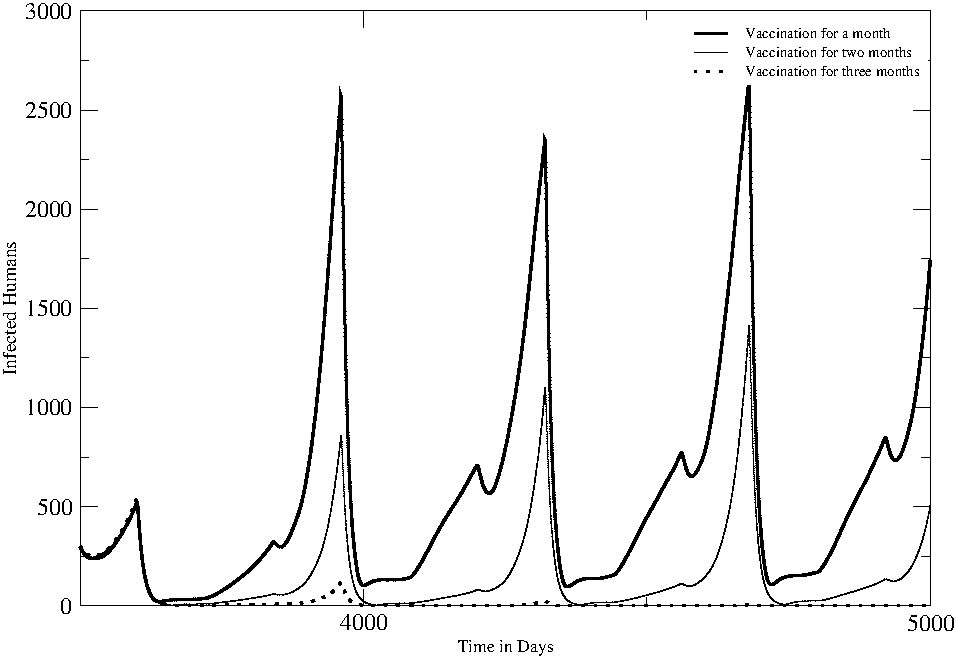}
 \caption{Evolution in time of dengue epidemic for different durations of the vaccination campaign.}
 \label{fig8}
\end{figure}

\begin{figure}[htb!]
 \centering
 \includegraphics[width=8.5cm]{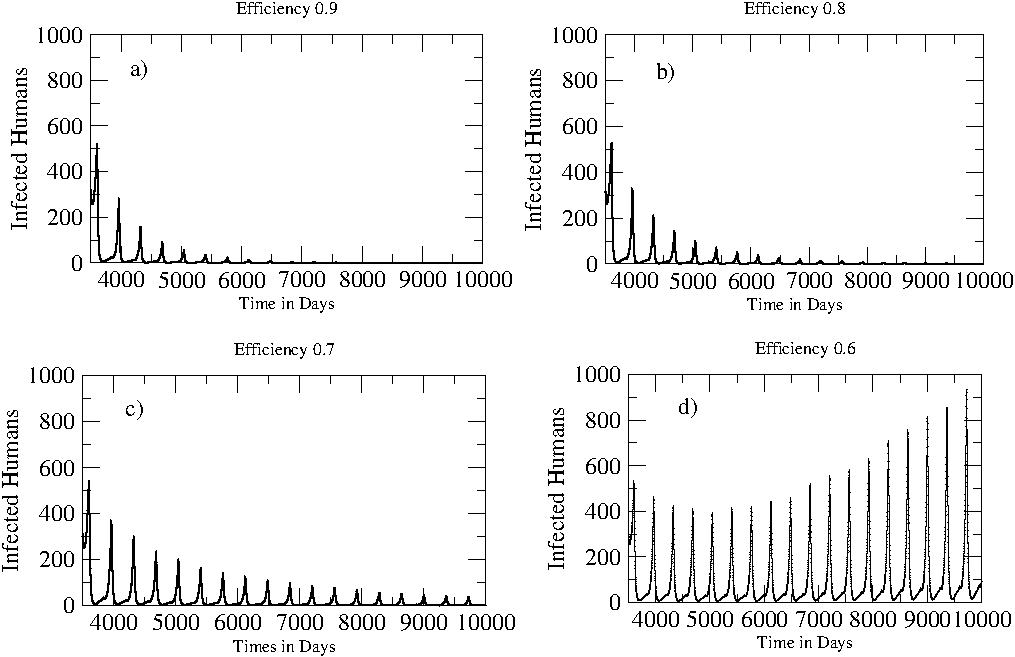}
 \caption{Comparison of different rates of efficiency for the vaccine against the dengue virus: (a) describes the efficiency of $90\%$, (b) $80\%$ efficiency, (c) $70\%$ efficiency, and (d) $60\%$ efficiency.}
 \label{fig9}
\end{figure}

With a vaccination rate fixed at $0.2$, and an efficiency maintained at $\xi=1$ the time during which the campaign should be maintained in order to eliminate the virus was estimated. Figure ($\ref{fig8}$) shows that a campaign active for three consecutive months eradicated the virus from the population. Furthermore, figure ($\ref{fig9}$) shows the evolution in time of the population infected humans for different values of $\xi$ for a campaign lasting three months. A vaccine that is $60\%$ efficient does not eliminate the virus and with $70\%$ there is a significant decrease in the number of cases over time. In order to eliminate the dengue fever of the human population, at least $80\%$ efficiency is required. Larger $\xi$ values, such as $90\%$, only differ in the time spent to eradicate the virus.

\subsection{Human population without exponential growth}

As variously mentioned, dengue is present in many countries, including Brazil and other South American countries, where the rate of population growth is not enough to reach 
levels as high as that of the African countries. According to a study by the Brazilian Institute of geography and statistics (IBGE\footnote{IBGE: Brazilian population ageing in 
rhythm accelerated}), the Brazilian population is decreasing its growth rate since the $60's$, when this rate was $3.04\%$. 
The IBGE estimates that this reduction in the growth rate still should last for many years reaching the so-called ``zero growth'' in $2039$ \cite{IBGE}. Hence, we also adjusted the birth rate of the human population to a value closer to the mortality rate. So, in these simulation the birth rate is now fixed in $0.000045$.

\begin{figure}[htb!]
 \centering
 \includegraphics[width=8.5cm]{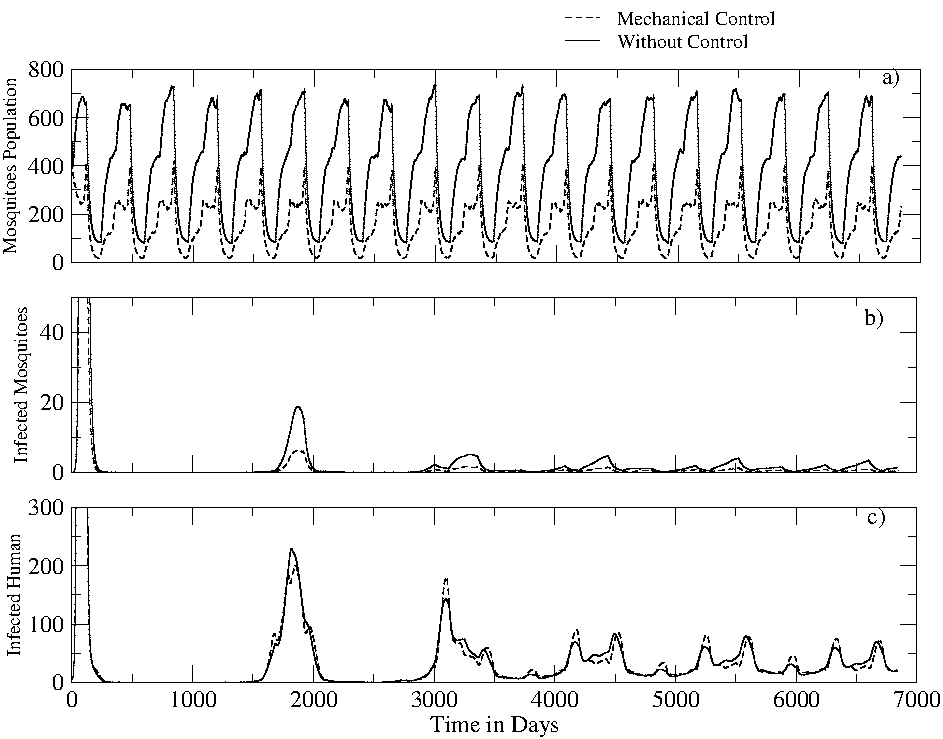}
 \caption{Comparison of mosquito (total and infected) and infected human infected population when no control is in effect and when it is done using the mechanical control throughout the year with birth rate $\mu_{d} = 0.000045$. (a) The total mosquitoes population, (b) the population of infected mosquitoes, (c) population of infected humans.}
 \label{fig10}
\end{figure}

Considering this new birth rate for the human population and the use of mechanical control throughout the year, an efficiency of $80\%$ was reached. However, as shown in 
figure ($\ref{fig10}$), the reduction of mosquitoes population does not decrease the infected human population. The behavior remains the same when the control was not used. 
We note that even with a low number of infected mosquitoes, the dengue virus still remains in the human population. This is possible because, as proven through histological 
studies, the dengue mosquito feeds on blood several times during a single cycle consequently, a single female can contaminate several humans in the same cycle \cite{Morrison}.

\begin{figure}[htb!]
 \centering
 \includegraphics[width=8.5cm]{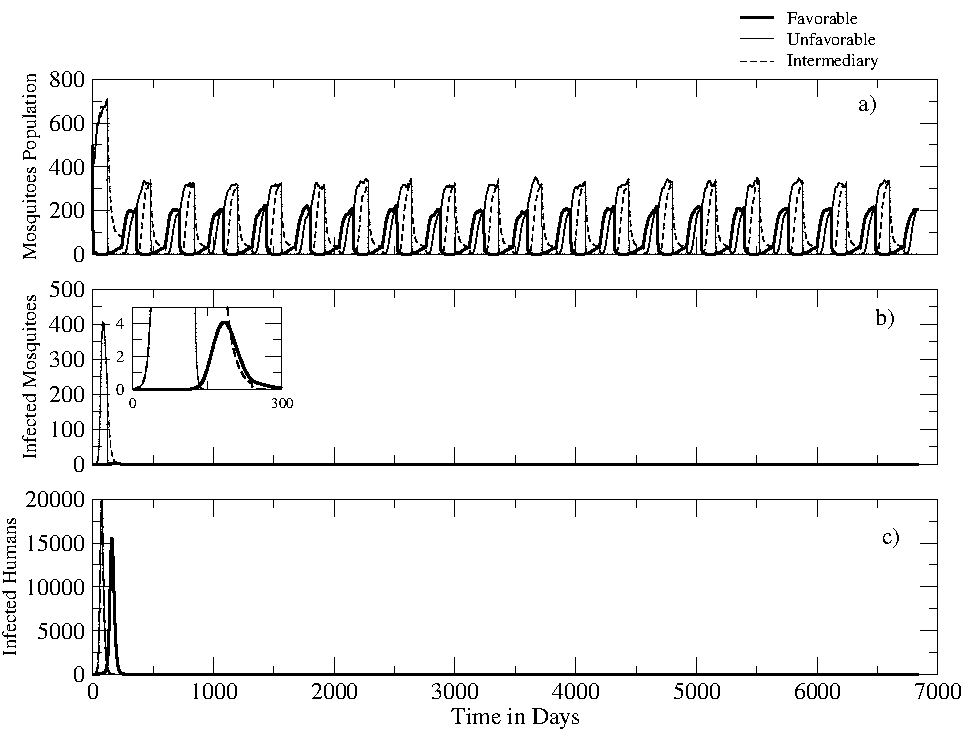}
 \caption{Evolution in time of mosquito and humans population with the use of chemical control and birth rate $\mu_{d}= 0.000045$ for the favorable, unfavorable and intermediate periods: (a) is the total population of mosquitoes, (b) is the population of infected mosquitoes, (c) is population of infected humans.}
 \label{fig11}
\end{figure}

Since only the mechanical control was enable to eliminate the dengue fever epidemic, a chemical control was applied in order to check its efficiency taking in to account this new birth rate. The system response to the use of insecticide and larvicidal applied in different periods of the year is shown in figure ($\ref{fig11}$).

Now we can notice that the chemical control is not efficient in eradicate the dengue viruses regardless of the period in which it is being applied, eliminating all 
infected mosquito in the system, but this does not reduce the levels significantly below the susceptible vector's population, as Luz \cite{Luz}. 
In the work done by Burattini and et. al, it was possible to evaluate the use of chemical control, which had a 
considerable impact to prevent the spread of dengue fever epidemic \cite{Burattini}.

But the persistence of \textit{Aedes aegypti} in the system, even if it is not infected, the dengue fever may reappear at any time if any infected person is introduced, and so the dengue epidemic can re-emerge in the human population, since the vector of the virus is still present.

It is known that the vector elimination campaigns by means of chemical agents often occurs in the favorable period of the year. Accordingly, we permed the chemical control in this period to see how the system respond to the emergence of infected people.

\begin{figure}[htb!]
 \centering
 \includegraphics[width=8.5cm]{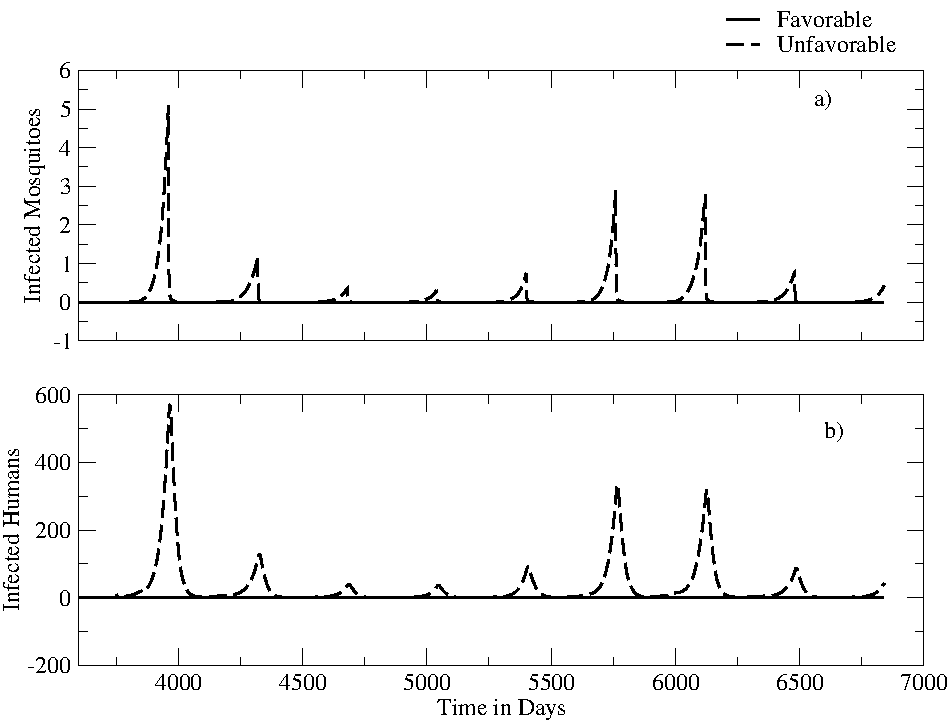}
 \caption{Evolution in time of populations of infected (a) mosquitoes (b) humans for a scenary in with $10$ infected humans appear in a system free of viruses. The analyses were made in the favorable and unfavorable periods.}
 \label{fig12}
\end{figure}

Figure ($\ref{fig12}$) shows the result of simulations in which $10$ individuals infected with the dengue virus appear at two different periods of the year. If these infected individuals come when chemical control is applied in the favorable period (continuous line), dengue does not reappear and the system remains free of the disease. On the contrary, if they are introduced in the unfavorable period, when there was no chemical control (dashed line), we have periodic outbreaks of the disease are observed. This suggests that a control campaign performed only in a period of the year may not be effective in eliminating dengue totally, since the virus can reappear through the emergence of infected people.

With the rate $\mu_{d} = 0.000045$ the human population does not grow exponentially, probably because 
the elimination of infected persons through the use of chemical control becomes possible. But even with the complete elimination of the 
population of infected vectors, it has not been possible to eliminate all the susceptible mosquito population in the system as shown in figure 
($\ref{fig12}$). The dengue fever may reappear at any time, unless there is a vaccine to protect the population against the virus.

\subsubsection{Vaccination campaign}

In this simulation the population does not exhibit a high growth rate and the vaccination campaign will be represented by the withdrawal of a fixed number 
of susceptible individuals ($\psi = 200$) by adding them into the recovered population. Thus, the vaccination rate is no longer proportional to the number 
of susceptible individuals, and is described by equations ($\ref{eqa}$) and ($\ref{eqb}$). In this case, a deemed efficiency $\xi=80\%$ for the vaccine is 
considered since it is the immunization threshold acceptable by health agencies. 
\begin{equation}
\label{eqa}
 \frac{d\,s(t)}{d\,t} = \dots - \psi\,\xi 
\end{equation}
\begin{equation}
\label{eqb}
\frac{d\,r(t)}{d\,t} = \dots + \psi\,\xi 
\end{equation}

\begin{figure}[htb!]
 \centering
 \includegraphics[width=8.5cm]{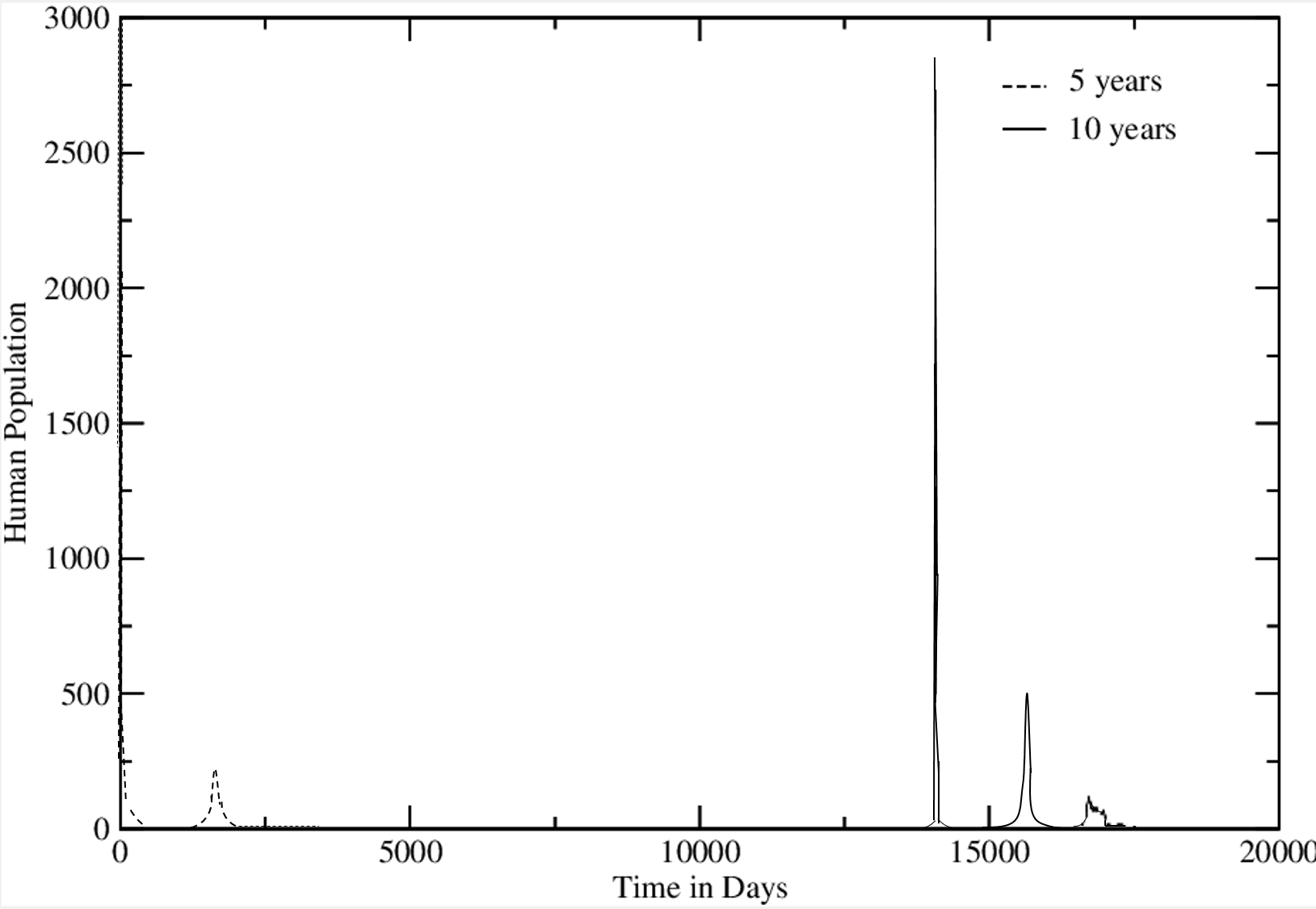}
 \caption{The vaccination campaign to immunize $200$ individuals per day with different periods of duration.}
 \label{fig13}
\end{figure}

Figure ($\ref{fig13}$) shows how the system behaves under immunization of $200$ people per day for $1$ month, with different times of duration of the campaign. It can be observed that, if the campaign remains for only $5$ consecutive years, dengue virus re-emerge in populations of mosquitoes and humans and the epidemic reappears. However, if the duration of the campaign is $10$ years, it is possible to eradicate the dengue virus from the human population for at least $30$ years, in a total simulations time fixed at $50$ years. To find out the time required for the permanent eradication of the virus from the system, more simulations considering longer simulation times, as well as the possibility of the emergence of other virus serotypes must be carried out.

Several studies show that a vaccine that protects the population from the dengue virus is about to be created. An attenuated tetravalent vaccine virus has been 
tested in order to characterize the interactions between the different serotypes of dengue virus, concluding that it is possible to create an effective vaccine against all 
four serotypes of the dengue virus \cite{Anderson}. However, an important point to be considered is whether live, attenuated virus vaccines do not offer risks to human health, 
since it is not clear that the presence of four types of strains into a single vaccine might increase the risk of developing severe forms of the disease \cite{Silva}.

\section{Conclusions}
\label{c}

Mathematical models are quite useful for observing the dynamics of infectious diseases such as the dengue fever. The creation of mathematical models that simulate the interaction between host-vector generates momentum in the system. Through the compartmental model proposed by Yang and Ferreira was used, it was possible to introduce small changes in order to test new control strategies. Testing the mechanical and chemical control verified the vaccination strategies for the human population.

The mechanical and chemical control were effective in reducing the population of Aedes aegypti but assured that the disease has been eliminated from the system. The control that used only the larvicidal had a very low contribution in this reduction, probably because the time interval between the implementation and its impact on the population of adults is very high, making this control to be less appropriate in controlling an epidemic of dengue fever. The mechanical control is much more efficient, which indicates that awareness campaigns are more useful than the use of controls through insecticide and larvicidal.

When considering a human population birth rate equal to $0.00042/day$, it was possible to simulate an exponential growth dynamics of this population. However, this consideration does not affect the effectiveness of mechanical control. In this case the application of mechanical control must be constant throughout the year. Once this control is active only in time after his application, it generates an increase in the number of cases in the other two periods, as there is no control being carried out.

With respect to the use of chemical control, there was an interference mosquito population in a yield of $21\%$ for the insecticide and larvicide only $5\%$. When the chemical and mechanical control were used together, there was an efficiency of $26\%$. In the latter case, even with greater efficiency, the use of these controls did not influence the dynamics of dengue in the human population, who remained infected. Importantly, the chemicals used in the simulations is fictitious and quite specific. Simulations with insecticide and larvicide with different duration should be conducted so that we can guarantee these results.

When the use of both controls was compared to situation where no control was accomplished, it resulted in an efficiency of $86\%$, compared to the situation where only the mechanical control was carried out in which the efficiency value fell to $26\%$, thus showing that the mechanical control is more efficient in the elimination of mosquito population than the chemical control even when applied throughout the year.

When the birth rate was changed to $0.000045/$\textit{day}, it was no longer possible to observe an exponential growth for the human population. In this case the efficiency of mechanical control remained at $80\%$. In the same way for the previous simulations, there was no significant reflection on human population infected. The chemical control can eliminate the infected population but the outbreaks do not end with a susceptible population. With $10$ individuals in the system infected with the dengue virus in the period in which the control is not being applied in the unfavorable time, periodic outbreaks of the disease occur.

The only way to contain the dengue epidemic, regardless of the form of population growth, is to create a vaccine to immunize the population against the virus. The population grew exponentially during the immunization vaccination campaign for a fraction proportional to a vaccination rate of individuals, $\nu = 0.2$. For this campaign to become really effective it is necessary to have a duration of three months and an efficiency rate of at least $80\%$.

It was possible to note the existence of annual cycles and a strong seasonality of epidemic cycles \cite{Nakhapakorn}. A major concern for the safety of live attenuated virus vaccines is the genetic stability. The use of molecular tools is required to ensure quality, safety and authenticity of genetic vaccine \cite{Rodrigues}. 

Already when the exponential growth is removed, the campaign is based on daily immunization $\psi = 200$, of people, with an efficiency of $80\%$. In this case it is necessary to maintain the campaign for at least $10$ years to ensure at least $30$ years of a system free of the disease.

Variable $\mu_{d}$ was introduced representing the portion of infected who die from complications related to dengue fever. The addition of this new factor caused the outbreaks of the disease might arise later, but as the simulation time is fixed at $50$ years this variable can have a major stake, because it may delay the outbreak so that it will not reappear during the time examined. However, field surveys are still necessary in order to better understand the mechanisms responsible for generating these dynamics. In addition, more simulations must be carried out in order to consider a longer time of observation, while factoring in the possibility of the emergence of other serotypes of the disease and biological ways of eliminating the vector.

Further studies are needed to understand the mechanisms responsible for generating the dynamics of dengue in order to learn how to fight it. The need for a vaccine is essential for the elimination of dengue viruses in human populations, thus it is expected that studies about immunization are satisfactory and achieve a positive result as soon as possible.





\end{document}